\documentclass[12pt]{article}
\usepackage{amsfonts}
\usepackage{amsmath}
\usepackage{amssymb}
\usepackage{mathptmx}
\usepackage{geometry}
\usepackage{graphicx}
\usepackage{hyperref}
\usepackage{cancel}
\usepackage{textcomp}
\usepackage{tikz}
\usepackage{bm}
  \usepackage{mathrsfs}
  \usepackage{filecontents}
\usepackage{times}
\usepackage{epsfig}
\usepackage{xcolor}
\usepackage{slashed}
\usepackage{booktabs}% build a table
\usepackage{latexsym}
\usepackage{verbatim}
\usepackage{extarrows}
\usepackage{multirow}
\usepackage{rotating}
\usepackage{colortbl}
\usepackage{indentfirst}
\usepackage{float}
\definecolor{mygray}{gray}{.9}
\definecolor{intnull}{RGB}{213,229,255}
\usepackage{arydshln}
\usepackage{diagbox}
\usepackage{makecell}
\usepackage{array}
\usepackage{cite}
\usepackage{caption}
\usepackage{tikz}
\usetikzlibrary{arrows,shapes,chains}
\usepackage{graphicx, subfig}
\begin{document}
\renewcommand{\thefootnote}{\fnsymbol{footnote}}
\baselineskip=16pt
\pagenumbering{arabic}
\vspace{1.0cm}
\begin{center}
{\Large\sf Can shadows connect black hole microstructures?}
\\[10pt]
\vspace{.5 cm}

{Xin-Chang Cai\footnote{E-mail address: caixc@mail.nankai.edu.cn} and Yan-Gang Miao\footnote{Corresponding author. E-mail address: miaoyg@nankai.edu.cn}}
\vspace{3mm}

{School of Physics, Nankai University, Tianjin 300071, China}

\vspace{4.0ex}

{\bf Abstract}
\end{center}

We investigate the relationship between shadow radii (photosphere radii) and black hole microstructures for a static spherically symmetric black hole and confirm their close connection.
As a concrete analysis, we take the Reissner-Nordstr$\mathrm{\ddot{o}}$m AdS black hole surrounded by perfect fluid dark matter (RN AdS black hole surrounded by PFDM) as an example. We calculate the Ruppeiner thermodynamic scalar curvature and shadow radius (photosphere radius) for the specific model. On the one hand, we find that the greater density of the perfect fluid dark matter makes black hole molecules more likely have attractive interactions. On the other hand, we provide the support to our general investigation that the shadow radii (photosphere radii) indeed connect the microstructures of this model.

\newpage

\section{Introduction}

Recently, the Event Horizon Telescope (EHT) collaboration has released~\cite{P4,P5} for the first time the shadow image of the supermassive black hole in the center of M8$7^{*}$ galaxy, which greatly stimulated our enthusiasm for the research of black hole shadows. A shadow is an observable quantity closely related to its corresponding black hole. Its essence is that the specific photons around a black hole collapse inward to form a dark area seen by a distant observer, which also means that the black hole shadow has a close connection to the dynamics of black holes. Through black hole shadows, one not only obtains~\cite{P6,P7,P8,P9,P10} the mass, spin, hair, and other information of black holes, but also knows~\cite{P11,PHXW,P12,P13} the distribution of matter around black holes. More importantly, the research in this direction opens a new window for us to study the strong gravitational region near black hole horizons. So far, there have been a large number of articles on black hole shadows under Einstein gravity and modified gravity theories, see, for instance, some literature~\cite{P44,P45,P46,P47,P48,P49,P50,P51,P52,P53,PX90,P54,P55,P56,P57,P58,P59,P71,P60,P61}.

As is known, the black hole thermodynamics~\cite{P14,P15,P16} is an important research field of black hole physics, especially when combined with AdS spacetimes, the black hole thermodynamics in extended phase spaces has achieved great progress, {\em e.g.}, the van der Waals-like phase transition occurs~\cite{P17} in the Reissner-Nordstr$\mathrm{ \ddot{o}}$m AdS black hole, the reentrant phase transition happens~\cite{P18} in high-dimensional rotating black holes, and so on. Black hole is not only a strong gravitational system, but also a thermodynamic one, so it is necessary to explore the relationship between its dynamics and thermodynamics. Recently, the relationships between dynamic characteristic quantities and thermodynamic ones have been studied~\cite{P19,P20,P21,P22,P23,P24,P25,P26,P27,P28}, such as photosphere radius, Hawking temperature, specific heat, and the minimum impact parameter used to describe the unstable circular orbital motion of photons around black holes, {\em et al}. The results show~\cite{P19,P20} that the unstable circular orbital motion can connect thermodynamic phase transitions of black holes. Especially for a static spherically symmetric black hole, the timelike or lightlike circular orbital motion of a test particle and its shadow indeed have a close connection~\cite{P22,P27,P28} to its thermodynamic phase transition.

The Ruppeiner thermodynamic geometry~\cite{P29,P30,P31,P32} is a useful tool for black hole thermodynamics. It provides helpful information about phase transitions of black holes, and also sheds~\cite{P33,P34,P35,P36,P37,P38,P39,P40,P41,P42,P43} some light on microstructures of black holes. Its most important physical quantity is the Ruppeiner thermodynamic scalar curvature. It has been shown~\cite{P30,P32,P38,P41,P42,P43} that this scalar curvature can take a positive, or a negative, or a vanishing value which corresponds to a repulsive, or an attractive, or no interaction among black hole molecules, respectively. As is known, thermodynamic phase transitions are not detectable directly, and neither are black hole microstructures. Therefore, it is necessary to associate a shadow radius with the Ruppeiner thermodynamic scalar curvature, so as to connect black hole microstructures with an observable -- the shadow radius. Furthermore, our investigations may provide some helps for understanding black hole microstructures by observing black hole shadows.

The paper is organized as follows. In Sec. 2, we investigate the relationship between shadow (photosphere) radii and black hole microstructures for a static spherically symmetric black hole. Then, we calculate in Sec. 3 the thermodynamic scalar curvature for the RN AdS black hole surrounded by PFDM. In Sec. 4, we study the relationship between the shadow radius and the Ruppeiner thermodynamic scalar curvature for the specific black hole system. Finally, we make a simple summary in Sec. 5. We use the units $c=G=k_{B}=\hbar=1$ and the sign convention $(-,+,+,+)$ throughout this paper.

\section{ Shadow and  black hole microstructure of static spherically symmetric black holes}

A static spherically symmetric black hole can be described by the line element,
\begin{equation}
\label{15}
ds^{2}=-f(r)dt^{2}+\frac{dr^{2}}{g(r)}+r^{2}\left(d\theta^{2}+\sin^{2}\theta d\varphi^{2}\right),
\end{equation}
where $f(r)$ and $g(r)$ are functions of the radial coordinate $r$.  We use the Hamilton-Jacobi equation to separate  the null geodesic equations in static spherically symmetric black hole spacetimes and derive the general expression of shadows of this class of black holes. The Hamilton-Jacobi equation takes~\cite{PSCH} the form,
\begin{equation}
\label{16}
\frac{\partial {\cal S}}{\partial \lambda }+{\cal H}=0, \qquad    {\cal H}=\frac{1}{2}g_{\mu \nu }p^{\mu }p^{\nu },
\end{equation}
where $\lambda$ is the affine parameter of null geodesics,  ${\cal S}$ the Jacobi action, ${\cal H}$ the  Hamiltonian, and the  momentum $p^{\mu }$ is defined by
\begin{equation}
\label{17}
p_{\mu }\equiv \frac{\partial {\cal S}}{\partial x^{\mu }}=g_{\mu \nu }\frac{\mathrm{d} x^{\nu }}{\mathrm{d}\lambda}.
\end{equation}

Due to the spherical symmetry of Eq.~(\ref{15}), without loss of generality, one can consider the photons moving on the equatorial plane with $\theta =\frac{\pi }{2}$.  So the the Jacobi action ${\cal S}$ can be decomposed into the following form,
\begin{equation}
\label{18}
{\cal S}=\frac{1}{2}m^{2}\lambda -Et+L\varphi +S_{r}(r),
\end{equation}
where $m$ is the mass of moving particles around black holes, for photons $m=0$, $E$ the energy of photons, $L$ the angular momentum of photons, and $S_{r}(r)$ a function of coordinate $r$. By substituting Eqs.~(\ref{15}), (\ref{17}) and (\ref{18}) into Eq.~(\ref{16}), one can get the following three equations describing the motion of photons on the equatorial plane,
\begin{eqnarray}
\frac{\mathrm{d} t}{\mathrm{d} \lambda }& =& \frac{E}{f(r)},\label{19} \\
\frac{\mathrm{d} r}{\mathrm{d} \lambda }&=& \pm \frac{\sqrt{g(r)}}{\sqrt{f(r)}}\frac{\sqrt{E^{2}r^{4}-L^{2}r^{2}f(r)}}{r^{2}},\label{20} \\
\frac{\mathrm{d} \varphi }{\mathrm{d} \lambda }&=& \frac{L}{r^{2} }.\label{22}
\end{eqnarray}
Combining Eq.~(\ref{20}) with Eq.~(\ref{22}), one has
\begin{equation}
\label{X1}
\frac{\mathrm{d} r}{\mathrm{d}\varphi }= \frac{\frac{\mathrm{d} r }{\mathrm{d} \lambda }}{\frac{\mathrm{d \varphi} }{\mathrm{d} \lambda }}=\pm r\sqrt{g(r)\left(\frac{r^{2}E^{2}}{L^{2}f(r)} -1\right )}.
\end{equation}
Again considering $\left.\frac{\mathrm{d} r}{\mathrm{d}\varphi }\right|_{r=\xi}=0$ at the turning point of photon orbits, one can rewrite Eq.~(\ref{X1}) as
\begin{equation}
\label{X2}
\frac{\mathrm{d} r}{\mathrm{d}\varphi }= \pm r\sqrt{g(r)\left(\frac{r^{2}f(\xi )}{\xi^{2}f(r)} -1\right )}.
\end{equation}
In addition, according to the equation satisfied by the effective potential $V(r)$,
\begin{equation}
\label{23}
\left (\frac{\mathrm{d} r}{\mathrm{d} \lambda }\right)^{2}+V(r)=0,
\end{equation}
one obtains
\begin{equation}
\label{24}
V(r)=g(r)\left(\frac{L^{2}}{r^{2}}-\frac{E^{2}}{f(r)}\right).
\end{equation}
Using the effective potential, one can determine~\cite{P27} the critical orbit radius of photons, {\em i.e.} the photosphere radius for a static spherically symmetric spacetime,
\begin{equation}
\label{25}
V(r)=0, \qquad \frac{\mathrm{d} V(r)}{\mathrm{d} r}=0,  \qquad  \frac{d^2 V(r)}{d r^2}<0.
\end{equation}
By substituting  Eq.~(\ref{24}) into Eq.~(\ref{25}), one can obtain the following relationships,
\begin{equation}
\label{26}
\frac{L^{2}}{E^{2}}=\frac{r^{2}_{\rm ps}}{f(r_{\rm ps})},
\end{equation}
\begin{equation}
\label{27}
r \frac{\mathrm{d} f(r)}{\mathrm{d} r}-2f(r)\bigg|_{r=r_{\rm ps}}=0,
\end{equation}
where $r_{\rm ps} $ is the radius of photospheres.

Considering that the light emitted from  a static observer at  position $r_{0}$  transmits into the past with an angle $\vartheta $ relative to the radial direction, one obtains~\cite{P53,PX90,P27}
\begin{equation}
\label{X3}
\cot\vartheta =\frac{\sqrt{g_{rr}}}{\sqrt{g_{\varphi \varphi }}}\frac{\mathrm{d} r}{\mathrm{d}\varphi }\bigg|_{r=r_{0} }=\pm \sqrt{\frac{{r_{0}}^{2}f(\xi )}{\xi^{2}f(r_{0})} -1}.
\end{equation}
When the relation $\sin^{2}\vartheta =\frac{1}{1+\cot^{2}\vartheta }$ is used, the above equation can be written as
\begin{equation}
\label{X4}
\sin^{2}\vartheta =\frac{{\xi }^{2}f(r_{0})}{{r_{0}}^{2}f(\xi)}.
\end{equation}
Therefore, the shadow radius of black holes observed by a static observer at position $r_{0}$ can be expressed~\cite{P13,P53,PX90,P27} by
\begin{equation}
\label{X5}
r_{\rm sh}\equiv r_{0}\sin\vartheta =\left.\xi \sqrt{\frac{f(r_{0})}{f(\xi )}}\right|_{\xi \rightarrow r_{\rm ps}}.
\end{equation}

In the Ruppeiner thermodynamic geometry, the most important physical quantity to describe the microstructures of black holes is the Ruppeiner thermodynamic scalar curvature $R$~\cite{P29,P30}. If the entropy representation is chosen~\cite{P64,P65,P72}, we obtain
\begin{equation}
\label{66}
\frac{\partial R}{\partial S}=\frac{\partial R}{\partial r_{\rm ps}}\frac{\mathrm{d} r_{\rm ps}}{\mathrm{d}r_{\rm H}}\frac{\mathrm{d}r_{\rm H} }{\mathrm{d}S},  \qquad
\frac{\partial R}{\partial S}=\frac{\partial R}{\partial r_{\rm sh}}\frac{\mathrm{d} r_{\rm sh}}{\mathrm{d}r_{\rm H}}\frac{\mathrm{d}r_{\rm H} }{\mathrm{d}S}.
\end{equation}
On the one hand, we know~\cite{P22,P27} the relations,
\begin{equation}
\label{39}
\frac{\mathrm{d} r_{\rm ps} }{\mathrm{d} r_{\rm H}}>0,  \qquad    \frac{\mathrm{d} r_{\rm sh} }{\mathrm{d} r_{\rm H}}>0.
\end{equation}
On the other hand, for a static spherically symmetric black hole, its
Bekenstein-Hawking entropy, $S=\frac{A}{4}$, where $A$ is the surface area, equals
\begin{equation}
\label{4}
S=\pi r^{2}_{\rm H},
\end{equation}
which means
$\frac{\mathrm{d}S }{\mathrm{d}r_{\rm H}}>0$ or $\frac{\mathrm{d}r_{\rm H} }{\mathrm{d}S}>0$. Consequently, the positivity or negativity of $\frac{\partial R}{\partial S}$ depends entirely on the positivity or negativity of $\frac{\partial R}{\partial r_{\rm ps}}$ or $\frac{\partial R}{\partial r_{\rm sh}}$.
%Therefore, Eq.~(\ref{66}) can be rewritten as
%\begin{equation}
%\label{67}
%\frac{\partial R}{\partial S}=\frac{1}{2\pi r_{\rm H}}\frac{\partial R}{\partial r_{\rm ps}}\frac{\mathrm{d}r_{ps} }{\mathrm{d} r_{\rm H}},  \qquad
%\frac{\partial R}{\partial S}=\frac{1}{2\pi r_{\rm H}}\frac{\partial R}{\partial r_{\rm sh}}\frac{\mathrm{d}r_{sh} }{\mathrm{d} r_{\rm H}}.
%\end{equation}
In other words, the conditions
\begin{equation}
\label{420}
\frac{\partial R}{\partial S}>0, \qquad   \frac{\partial R}{\partial S}=0, \qquad   \frac{\partial R}{\partial S}<0,
\end{equation}
correspond to
\begin{equation}
\label{42}
\frac{\partial R}{\partial r_{\rm ps}}>0, \qquad   \frac{\partial R}{\partial r_{\rm ps}}=0, \qquad   \frac{\partial R}{\partial r_{\rm ps}}<0,
\end{equation}
or equivalently to
\begin{equation}
\label{421}
\frac{\partial R}{\partial r_{\rm sh}}>0, \qquad   \frac{\partial R}{\partial r_{\rm sh}}=0, \qquad   \frac{\partial R}{\partial r_{\rm sh}}<0.
\end{equation}
We thus establish the relationship between the shadow (photosphere) radius and the microstructures of a static spherically symmetric black hole. Next, we take the RN AdS black hole surrounded by PFDM as an example to compute  its Ruppeiner thermodynamic scalar curvature and analyze the connection between the scalar curvature and the shadow (photosphere) radius.

\section{RN AdS black hole surrounded by  PFDM and  its Ruppeiner thermodynamic geometry  }

The line element of the RN AdS black hole surrounded by  PFDM is given~\cite{P66,P67,P68} by
\begin{equation}
\label{1}
ds^{2}=-f(r)dt^{2}+\frac{dr^{2}}{f(r)}+r^{2}\left(d\theta^{2}+\sin^{2}\theta d\varphi^{2}\right),
\end{equation}
with
\begin{equation}
\label{2}
f(r)=1-\frac{2M}{r}+\frac{Q^{2}}{r^{2}}-\frac{\Lambda r^{2}}{3}+\frac{\alpha }{r}\textrm{ln}\left(\frac{r}{|\alpha|}\right).
\end{equation}
Here, $M$ is the black hole mass, $Q$ its charge, $\Lambda$ the negative cosmological constant, and $\alpha $ the intensity parameter related to the perfect fluid dark matter.
For the case of $\alpha=0 $, the above metric turns back to that of the RN AdS black hole, which means no perfect fluid dark matter outside the RN AdS black hole.

In the extended phase space including the cosmological constant, the mass $M$ regarded as the black hole enthalpy $H$ can be expressed~\cite{P68} by
\begin{eqnarray}
\label{3}
M=H&=&\frac{r_{\rm H}}{2}+\frac{Q^{2}}{2r_{\rm H}}+\frac{4\pi P}{3}r^{3}_{\rm H}+\frac{\alpha }{2}\textrm{ln}\left(\frac{r_{\rm H}}{|\alpha |}\right)\nonumber \\
&=&\frac{\sqrt{S}}{2\sqrt{\pi }}+\frac{\sqrt{\pi }Q^{2}}{2\sqrt{S}}+\frac{4PS^{3/2}}{3\sqrt{\pi }}+\frac{\alpha }{2}\textrm{ln}\left(\frac{\sqrt{S}}{\sqrt{\pi }|\alpha |}\right),
\end{eqnarray}
and the Hawking temperature by
\begin{eqnarray}
\label{5}
T&=&\frac{1}{4\pi r_{\rm H}}-\frac{Q^{2}}{4\pi r^{3}_{\rm H} }+2Pr_{\rm H}+\frac{\alpha }{4\pi r^{2}_{\rm H}}\nonumber \\
&=&\frac{1}{4\sqrt{\pi S}}-\frac{\sqrt{\pi}Q^{2}}{4S^{{3}/{2}}}+\frac{2\sqrt{S}P}{\sqrt{\pi }}+\frac{\alpha }{4S},
\end{eqnarray}
where $f(r_{\rm H})=0$, $P=-\frac{\Lambda}{8\pi}$, and Eq.~(\ref{4}) have been considered.
%Using $T=\frac{1}{4\pi}f'(r_{\rm H})$, we calculate
In addition, the thermodynamic volume $V$ conjugated to the pressure $P$, the electric potential $\Phi$ conjugated to the charge $Q$, and the extensive quantity $\Psi$ conjugated to the intensity parameter $\alpha$ are defined respectively by
\begin{eqnarray}
V &\equiv& \frac{\partial H}{\partial P}=\frac{4\pi }{3}r^{3}_{\rm H},\label{7} \\
%\end{equation}
%\begin{equation}
\Phi &\equiv& \frac{\partial H}{\partial Q}=\frac{Q}{r_{\rm H}},\label{8} \\
%\end{equation}
%\begin{equation}
\Psi &\equiv& \frac{\partial H}{\partial \alpha}=\frac{1}{2}\left[\textrm{ln}\left(\frac{r_{\rm H}}{|\alpha |}\right)-1\right].
\end{eqnarray}
Correspondingly, for the RN AdS black hole surrounded by PFDM, the first law of thermodynamics and the Smarr relation take the forms in terms of the thermodynamic variables defined above, respectively,
\begin{equation}
\label{9}
dM=dH=TdS+\Phi dQ+VdP+\Psi d \alpha ,
\end{equation}
and
\begin{equation}
\label{10}
M=H=2TS+\Phi Q-2VP+\Psi \alpha.
\end{equation}

Now we follow the Ruppeiner thermodynamic geometry to study the RN AdS black hole surrounded by PFDM. Here we use the enthalpy and Helmholtz free energy representations, that is, we calculate the Ruppeiner thermodynamic scalar curvature $R$ in $(S, P)$ and $(T, V)$ planes, respectively. The Ruppeiner line element of this black hole spacetime in the $(S, P)$ plane is~\cite{P64,P65,P72} as follows:
\begin{equation}
\label{11}
ds^{2}_{R}=\frac{1}{C_{P}}dS^{2}+\frac{2}{T}\left (\frac{\partial T}{\partial P}\right)_{S} dSdP,
%+\frac{1}{T}\left(\frac{\partial V}{\partial P}\right)_{S}dP^{2},
\end{equation}
where $C_{P}\equiv T\left(\frac{\partial S}{\partial T} \right)_{P}$ and the relation, $\left(\frac{\partial V}{\partial P}\right)_{S}=0$, has been used because the thermodynamic volume $V$ is independent of the presure $P$ under a fixed entropy $S$. Similarly, the Ruppeiner line element  in the $(T,V)$ plane is given~\cite{P64,P65,P72} by
\begin{equation}
\label{12}
ds^{2}_{R}=\frac{2}{T}\left(\frac{\partial P}{\partial T} \right )_{V}dTdV+\frac{1}{T} \left( \frac{\partial P}{\partial V}\right)_{T}dV^{2},
\end{equation}
where $C_{V}\equiv T\left(\frac{\partial S}{\partial T} \right)_{V}=0$ has been considered.

Substituting Eqs.~(\ref{4}), (\ref{5}), and (\ref{7}) into the line elements Eqs.~(\ref{11}) and (\ref{12}), we obtain the Ruppeiner thermodynamic scalar curvatures in the $(S, P)$ and $(T, V)$ planes, respectively,
\begin{equation}
\label{13}
R_{SP}=-\frac{2(S-2\pi Q^{2})+3\sqrt{\pi S}\alpha }{2S(8PS^{2}+S-\pi Q^{2}+\sqrt{\pi S}\alpha )},
\end{equation}
and
\begin{equation}
\label{14}
R_{VT}=-\frac{6\sqrt[3]{\pi}V^{{2}/{3}}-8\sqrt[3]{6}\pi Q^{2}+3\sqrt[3]{(6\pi)^{2}}\alpha V^{{1}/{3}}}{18\sqrt[3]{\pi^{4}}V^{{5}/{3}}T}.
\end{equation}
We can easily check that these two scalar curvatures are equal to each other, that is, $R_{SP}=R_{VT}$. It should be noted that the Ruppeiner thermodynamic scalar curvature diverges when the black hole goes to its extreme configuration. In addition, for the case of $\alpha=0$, Eqs.~(\ref{13}) and (\ref{14}) will return to the thermodynamic curvature of RN AdS black holes~\cite{P65}.

In Fig. 1, based on Eq.~(\ref{13}) we depict the relation of the Ruppeiner thermodynamic scalar curvature $R$ with respect to the entropy $S$ for different values of parameter $\alpha$ at a constant pressure $P$ and a constant charge $Q$. In Fig. 2, based on Eq.~(\ref{14}) we depict the relation of the Ruppeiner thermodynamic scalar curvature $R$ with respect to the thermodynamic volume $V$ for different values of parameter $\alpha$ at a constant Hawking temperature $T$ and a constant charge $Q$. We notice that the profile of $R$ versus $S$ is similar to that of $R$ versus $V$. This is not strange because $S$ is proportional to the square of $r_{\rm H}$ and $V$ to the cubic of $r_{\rm H}$, see Eqs.~(\ref{4}) and (\ref{7}). From the two figures, we find that, on the one hand, the black hole molecules of the RN AdS black hole surrounded by PFDM may have a repulsive interaction ($R>0$), or an attractive interaction ($R<0$), or even no interaction ($R=0$), depending on the regions of the black hole's intrinsic parameters; and on the other hand, the increase of $\alpha$ makes the curves move downward, which indicates that the greater density of the perfect fluid dark matter makes the black hole molecules more likely have attractive interactions. %In addition, we find that the effect of $\alpha$ on interactions among the molecules of the RN AdS black hole is different from that of $\alpha$ related to the quintessence field in Ref.~\cite{P69}.

\begin{figure}
\centering
\begin{minipage}[t]{0.6\linewidth}
\centering
\includegraphics[width=100mm]{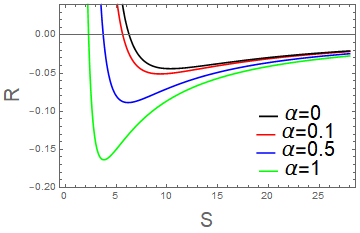}
\caption*{Figure 1. Graph of the Ruppeiner thermodynamic scalar curvature $R$ with respect to the entropy $S$ for different values of parameter $\alpha$ at a constant pressure $P$ and a constant charge $Q$. Here we choose $P=0.002$ and $Q=1$.}
\label{fig1}
\end{minipage}
\end{figure}

\begin{figure}
\centering
\begin{minipage}[t]{0.6\linewidth}
\centering
\includegraphics[width=100mm]{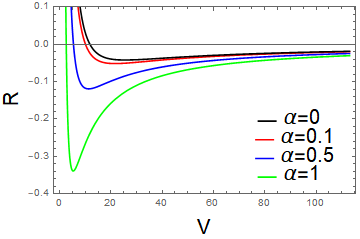}
\caption*{Figure 2. Graph of the Ruppeiner thermodynamic scalar curvature $R$  with respect to the thermodynamic volume $V$ for different values of parameter $\alpha$ at a constant Hawking temperature $T$ and a constant charge $Q$. Here we choose $T=0.04$ and $Q=1$.}
\label{fig2}
\end{minipage}
\end{figure}

\section{Shadow radius and Ruppeiner thermodynamic scalar curvature of the RN AdS black hole surrounded by PFDM}

By substituting Eq.~(\ref{2}) into Eqs.~(\ref{27}) and ~(\ref{X5}), we derive the shadow radius of the RN AdS black hole surrounded by PFDM observed by a static observer at position $r_{0}$,
\begin{equation}
\label{43}
r_{\rm sh}=r_{\rm ps}\sqrt{\frac{f(r_{0})}{f(r_{\rm ps})}},
\end{equation}
where the photosphere radius $r_{\rm ps}$ satisfies the following equation,
\begin{equation}
\label{44}
\frac{4Q^{2}}{r_{\rm ps}}+2r_{\rm ps}+3\alpha\, \textrm{ln}\left(\frac{r_{\rm ps}}{ | \alpha|}\right)-6M-\alpha =0.
\end{equation}
For the case of $\alpha=0$, {\em i.e.} the RN AdS black hole, we obtain the analytic expression of the photosphere radius in terms of $M$ and $Q$,
\begin{equation}
\label{45}
{\tilde r}_{\rm ps}=\frac{1}{2}\left(3M+\sqrt{9M^{2}-8Q^{2}}\right),
\end{equation}
and that of the shadow radius in terms of $M$, $Q$, and $r_0$,
\begin{equation}
\label{46}
{\tilde r}_{\rm sh}=\frac{\left(\sqrt{9 M^2-8 Q^2}+3 M\right)^2 \sqrt{-2 M r_0+\frac{8}{3} \pi  P r_0^4+Q^2+r_0^2}}{4 r_0 \sqrt{\frac{1}{6} \pi  P \left(\sqrt{9 M^2-8 Q^2}+3 M\right)^4+\frac{1}{4} \left(\sqrt{9 M^2-8 Q^2}+3 M\right)^2-M \left(\sqrt{9 M^2-8 Q^2}+3 M\right)+Q^2}}.
\end{equation}

By substituting Eq.~(\ref{3}) into Eqs.~(\ref{43}) and ~(\ref{44}), we cannot derive the analytic formulas of the photosphere radius and the shadow radius but can express the two radii as functions of $r_{\rm H}$, $Q$, $P$, and $\alpha$,
\begin{equation}
\label{D1}
r_{\rm ps}=r_{\rm ps}(r_{\rm H}, Q, P, \alpha),
\end{equation}
and
\begin{equation}
\label{47}
r_{\rm sh}=r_{\rm sh}(r_{\rm H}, Q, P, \alpha, r_0).
\end{equation}
For the case of $\alpha=0$, we can solve from Eqs.~(\ref{D1}) and (\ref{47}) the analytic expressions of the photosphere radius and the shadow radius for the RN AdS black hole,
\begin{equation}
\label{D2}
{\hat r}_{\rm ps}=\frac{\sqrt{\left(8 \pi  P r_{\rm H}^4+3 r_{\rm H}^2+3 Q^2\right){}^2-32 Q^2 r_{\rm H}^2}+8 \pi  P r_{\rm H}^4+3 r_{\rm H}^2+3 Q^2}{4 r_{\rm H}},
\end{equation}
and
\begin{equation}
\label{48}
{\hat r}_{\rm sh}=\frac{\left(\sqrt{B+8 Q^2}+\sqrt{B}\right)^2 \sqrt{-2 r_0 \sqrt{B+8 Q^2}+8 \pi  P r_0^4+3 Q^2+3 r_0^2}}{2 \sqrt{2} r_0 \sqrt{8 \pi  B^2 P+8 \pi  B P \left(\sqrt{B \left(B+8 Q^2\right)}+8 Q^2\right)+\left(\sqrt{B \left(B+8 Q^2\right)}+2 Q^2\right) \left(32 \pi  P Q^2+1\right)+B}},
\end{equation}
with
\begin{equation}
B\equiv \frac{\left(8 \pi  P r_{\rm H}^4+3 r_{\rm H}^2+3 Q^2\right){}^2-32 Q^2 r_{\rm H}^2}{4 r_{\rm H}^2}.
\end{equation}

Substituting Eq.~(\ref{4}) into Eq.~(\ref{13}), we compute the Ruppeiner thermodynamic scalar curvature as a function of $r_{\rm H}$, $Q$, $P$, and $\alpha$,
\begin{equation}
\label{49}
R=R(r_{\rm H}, Q, P, \alpha)=\frac{4 Q^2-2 r_{\rm H}^2-3 \alpha  r_{\rm H}}{2 \pi  r_{\rm H}^2 \left(8 \pi  P r_{\rm H}^4+r_{\rm H}^2-Q^2+\alpha  r_{\rm H}\right)}.
\end{equation}
For the case of $\alpha=0$, we obtain the Ruppeiner thermodynamic scalar curvature for the RN AdS black hole,
\begin{equation}
\label{50}
{\hat R}=\frac{2 Q^2-r_{\rm H}^2}{\pi  r_{\rm H}^2 \left(8 \pi  P r_{\rm H}^4+r_{\rm H}^2-Q^2\right)}.
\end{equation}

According to Eqs.~(\ref{D1}), (\ref{47}) and (\ref{49}), we deduce that there is a correspondence between  the Ruppeiner thermodynamic scalar curvature $R$ and the photosphere radius $r_{\rm ps}$ or between $R$ and the shadow radius $r_{\rm sh}$ for the RN AdS black hole surrounded by PFDM if charge $Q$, pressure $P$, and parameter $\alpha$ are fixed, which cannot be solved analytically but can be expressed as
\begin{equation}
\label{D3}
R=\left.R(r_{\rm ps})\right|_{Q=const., \,P=const., \,\alpha =const.},
\end{equation}
or
\begin{equation}
\label{51}
R=\left.R(r_{\rm sh})\right|_{Q=const., \,P=const., \,\alpha =const.}.
\end{equation}

In Fig. 3 for the RN AdS black hole surrounded by PFDM, based on Eqs.~(\ref{D1}) and (\ref{47}) we depict the relation of the photosphere radius $r_{\rm ps}$ with respect to the horizon radius  $r_{\rm H}$ and also the relation of the shadow radius $r_{\rm sh}$ with respect to the horizon radius $r_{\rm H}$  for different values of parameter $\alpha$ at a constant pressure $P$ and a constant charge $Q$. From this figure, we can see that the photosphere radius $r_{\rm ps}$ and the shadow radius $r_{\rm sh}$ increase monotonously with the increase of the horizon radius $r_{\rm H}$, which is completely consistent with Eq.~(\ref{39}) --- a general property of a static spherically symmetric black hole.

\begin{figure}
		\centering
		\begin{minipage}{.6\textwidth}
			\centering
			\includegraphics[width=100mm]{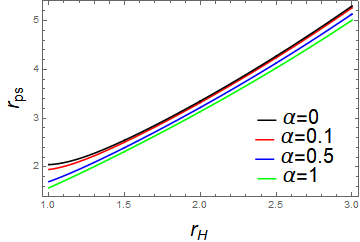}
		\end{minipage}
\vskip 10mm
		\begin{minipage}{.6\textwidth}
			\centering
			\includegraphics[width=100mm]{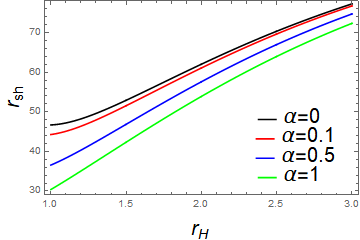}
		\caption*{Figure 3. Graphs of the photosphere radius $r_{\rm ps}$ and the shadow radius $r_{\rm sh}$ with respect to the horizon radius $r_{\rm H}$ for different values of parameter $\alpha$ at a constant pressure $P$ and a constant charge $Q$. Here we choose $P=0.002$, $Q=1$, and $r_{0}=100$.}
\label{figure4}
\end{minipage}
\end{figure}

In Figs. 4 and 5, based on Eqs.~(\ref{D3}) and (\ref{51}) we turn to the relation of the Ruppeiner thermodynamic scalar curvature $R$ with respect to the photosphere radius $r_{\rm ps}$ and also the relation of $R$ with respect to the shadow radius $r_{\rm sh}$ for different values of parameter $\alpha$ at a constant pressure $P$ and a constant charge $Q$. By comparing Fig. 1 with Figs. 4 and 5, we can clearly see that they have similar profiles, which indicates that the photosphere radius $r_{\rm ps}$ and the shadow radius $r_{\rm sh}$ can connect the microstructures of black holes. Therefore, the shadow radius is indeed a good observable that has a close connection to the black hole microstructures.

\begin{figure}
\centering
\begin{minipage}[t]{0.6\linewidth}
\centering
\includegraphics[width=100mm]{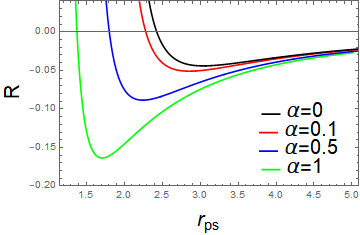}
\caption*{Figure 4. Graph of the Ruppeiner thermodynamic scalar curvature $R$ with respect to the photosphere radius $r_{\rm ps}$ for different values of parameter $\alpha$ at a constant pressure $P$ and a constant charge $Q$. Here we choose $P=0.002$ and $Q=1$.}
\label{fig1}
\end{minipage}
\end{figure}

\begin{figure}
\centering
\begin{minipage}[t]{0.6\linewidth}
\centering
\includegraphics[width=100mm]{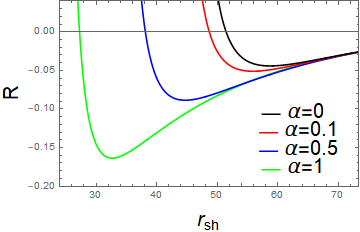}
\caption*{Figure 5. Graph of the Ruppeiner thermodynamic scalar curvature $R$ with respect to the shadow radius $r_{\rm sh}$ for different values of parameter $\alpha$ at a constant pressure $P$ and a constant charge $Q$. Here we choose $P=0.002$, $Q=1$, and $r_{0}=100$.}
\label{fig1}
\end{minipage}
\end{figure}

\section{Conclusion }

In this paper, we investigate for a static spherically symmetric black hole the relationship between the shadow radius (the photosphere radius) and black hole microstructures, and confirm their close connection. As a concrete analysis, we take the RN AdS black hole surrounded by PFDM as an example. We calculate the Ruppeiner thermodynamic scalar curvature for the specific model. On the one hand, we find that the black hole molecules may have a repulsive interaction, or an attractive interaction, or even no interaction, depending on the regions of the black hole's intrinsic parameters. On the other hand, we discover that the greater density of the the perfect fluid dark matter makes the black hole molecules more likely become attractive interactions. In addition, we observe from Figs. 1, 4, and 5 that the portrait of the Ruppeiner thermodynamic scalar curvature $R$ with respect to the shadow radius $r_{\rm sh}$ (the photosphere radius $r_{\rm ps}$) is similar to that of the thermodynamic curvature with respect to the entropy $S $ (the thermodynamic volume $V$). We thus establish the relationship between the shadow (photosphere) radius and black hole microstructures for the RN AdS black hole surrounded by PFDM, providing the support to our general investigation for a static spherically symmetric black hole.

An interesting issue is to explore whether the shadow (photosphere) radius of rotating AdS black holes has a close connection to black hole microstructures, which will be considered in our future work.

\section*{Acknowledgments}

The authors would like to thank Z.-M. Xu  and M. Zhang for helpful discussions.
This work was supported in part by the National Natural Science Foundation of China under Grant No. 11675081.

%The authors would like to thank the anonymous referee for the helpful comments that improve this work greatly.

%\newpage

\end{document}